\documentclass[letterpaper,times]{IONconf}
\usepackage{amsmath}
\usepackage{amssymb}
\usepackage{bm}

\usepackage[pdftex]{graphicx}
\usepackage[justification=centering]{caption}
\usepackage{subcaption}
\graphicspath{{assets/}}
\usepackage{algorithmic}

\usepackage{array}

\usepackage{url}

\usepackage[%
backend=biber,
style=apa,
]{biblatex}

\usepackage[hidelinks]{hyperref}

\usepackage{orcidlink}
\usepackage{appendix}

\title{Assessment of cryptographic approaches for a quantum-resistant Galileo OSNMA} 

\author{
    Javier Junquera-Sánchez \orcidlink{0000-0002-4597-6539}. {National Institute for Aerospace Technology (INTA)}
    \vspace{1mm} \\%
    Carlos Hernando-Ramiro \orcidlink{0000-0003-1798-1303}. {National Institute for Aerospace Technology (INTA)}
    \vspace{1mm} \\%
    Oscar Gamallo-Palomares \orcidlink{0009-0008-9734-5561}. {National Institute for Aerospace Technology (INTA)}
    \vspace{1mm} \\%
    José-Antonio Gómez-Sánchez \orcidlink{0009-0009-2351-7823}. {National Institute for Aerospace Technology (INTA)}
}

\ifdefined\isPreprint
    \let\oldMaketitle\maketitle
    \renewcommand{\maketitle}{
    \fbox{%
        \parbox{\textwidth}{
            \centering
            Submitted to NAVIGATION, Journal of the Institute of Navigation. \\
            
            See \href{https://navi.ion.org}{navi.ion.org}
            }%
        }
        \oldMaketitle
    }
\fi

\begin{document}

\maketitle

\fbox{%
    \parbox{\textwidth}{
        \centering
       Published in NAVIGATION: Journal of the Institute of Navigation Jun 2024, 71 (2) navi.648; \\
       \textbf{DOI: 10.33012/navi.648} \\
       See \url{https://navi.ion.org/content/71/2/navi.648}
    } %
}

\section*{Correspondence}

\newcommand{\email}[1]{\href{mailto:#1}{\texttt{#1}}}

Javier Junquera-Sánchez, National Institute for Aerospace Technology, Ajalvir road, km 4, 28850 Torrejón de Ardoz, Madrid, Spain.
Email: \email{jjunsan@inta.es}




\section*{Abstract}

As time goes on, quantum computing becomes more of a reality, bringing several cybersecurity challenges. Modern cryptography is based on the computational complexity of specific mathematical problems, but as new quantum-based computers appear, classical methods might not be enough to secure communications. In this paper, we analyse the state of the Galileo Open Service Navigation Message Authentication (OSNMA) to overcome these new threats. This analysis and its assessment have been performed using OSNMA documentation, reviewing the available Post Quantum Cryptography (PQC) algorithms competing in the National Institute of Standards and Technology (NIST) standardization process, and studying the possibility of its implementation in the Galileo service. The main barrier to adopting the PQC approaches is the size of both the signature and the key. The analysis shows that OSNMA is not yet prepared to face the quantum threat, and a significant change would be required. This work concludes by assessing different transitory countermeasures that can be implemented to sustain the system's integrity in the short term.


\section{INTRODUCTION}
\label{sec:s1}

Galileo is the global navigation satellite system (GNSS) developed by the European Union (EU) to provide positioning and timing information worldwide. Galileo provides Europe sovereignty over its navigation capabilities, avoiding any dependence on the other available GNSS, namely GPS (USA), GLONASS (Russia) and BeiDou (China) (\cite{langley_introduction_2017}). Galileo is the only GNSS under civil control, and, in addition to its well-known Open Service (OS), public and free of charge, offers other innovative services. Regarding security, the most relevant service is the Open Service Navigation Message Authentication (OSNMA), as currently, no other GNSS service offers a similar capability. OSNMA is freely accessible to all Galileo users and protects them against spoofing attacks intending to fake the real position and time of the receiver. For this purpose, OSNMA authenticates the data transmitted in the Galileo OS navigation messages (I/NAV), ensuring that it comes from the system itself and has not been modified. This is accomplished by transmitting authentication-specific data in previously reserved fields of the I/NAV message.

The OSNMA service is based on two building blocks:

\begin{enumerate}
    \item   Time Efficient Stream Loss-Tolerant Authentication (\cite{perrig_timed_2005}) (TESLA): a protocol that allows the authentication of messages using symmetric keys, as long as they do not become disclosed in a defined period of time.
    \item   Public-key cryptography: for secure authentication of the TESLA keys, establishing the root of trust over the long-term.
\end{enumerate}

Even though the trade-offs of TESLA have been discussed in other works (\cite{neish_parameter_2018}), its security depends on the correct selection of parameters (\cite{fernandez-hernandez_analysis_2021}), and it is required to be bootstrapped using Public Key Cryptography (PKC) (\cite{fries_bootstrapping_2006}). In this context, the reliability of OSNMA is crucial, as there are plans for using it to support other Galileo services, such as the Assisted Commercial Authentication Service (ACAS) (\cite{fernandez-hernandez_semi-assisted_2022, fernandez-hernandez_semi-assisted_2023}).

However, with the evolution of quantum computers capable of running algorithms, like Shor’s (\cite{shor_algorithms_1994}), the ``classical'' public-key cryptosystems (e.g. relying on mathematical problems such as discrete logarithm, integer factorization, etc.) will not be secure anymore. In particular, if an eventual quantum computer is built with the capability to implement these algorithms, the public key cryptosystem used in OSNMA (i.e. elliptic curves) would become vulnerable (\cite{chen_satellite_2023}), putting at risk the whole system (\cite{eledlebi_empirical_2022}).

All the relying symmetric cryptography elements would be weakened by other quantum computing algorithms, like Grover's (\cite{grover_fast_1996}), capable of searching preimages in an asymptotic square root time. This is, for instance, the case of MAC functions whose primitive security (i.e. hash functions) could be degraded to half of their current strength (\cite{10014648}).

To face the threats of quantum computing, several entities are developing cryptographic algorithms resistant against quantum adversaries. The most recognized standardization process is the Post-Quantum Cryptography (PQC) competition hosted by the National Institute of Standards and Technology (NIST) (\cite{computer_security_division_post-quantum_2017}). There are also other projects addressing the transition to Quantum-Safe Cryptography (\cite{dahmen-lhuissier_quantum-safe_nodate}).

According to Mosca’s theorem (\cite{mosca_cybersecurity_2018}), to ensure the security of information systems, it must be taken into account, not only the time until the maturity of quantum computers but also the period that the update of current information systems will take. It is also important to remark that, today’s confidentiality of information, will also be at risk against the ``store-now, decrypt-later'' strategy (\cite{joseph_transitioning_2022}). The development of Post-Quantum capabilities is essential for the prevention of vulnerabilities in communication networks (\cite{lopez_new_2022}), but this transition is not trivial and arises conflicts over information systems’ requisites: e.g. resources consumption, size of cryptographic data (\cite{westerbaan_sizing_2021}), etc. The security of the new-fangled PQC algorithms has not been tested enough yet, and the changes required to be implemented can collide with current standards.

Furthermore, some applications, such as timestamp digital signatures, or certificates emitted by Certificate Authorities (CA), should remain trustable for decades. However, the keys employed for these purposes face the risk to become insecure within this period: e.g. the Global Sign Root CA certificate must remain valid for 30 years, from 1/9/1998 to 28/1/2028, and there even exist TLS root certificates valid up to 2060 (\cite{inc_globalsign_2022}).
The EU has evidenced that space-based development solutions have reached a high level of implantation in society, and their degradation could derive in a threat to civil security. Even though it could seem remote, developing security strategies to address the implications of space-related risks is strategic to achieve resiliency (\cite{european_commision_eu_nodate}).
This research aims to provide an assessment of the options to implement PQC in Galileo’s OSNMA service, setting foundations to adopt the findings in other EU Space Projects. 

\subsection{Research questions}
\label{sub:rqs}

The present research aims to answer the following questions:

\begin{enumerate}
    \item Is it possible to make OSNMA quantum-resistant assuming its current design?
    \begin{enumerate}
        \item What is the maximum size available for a public key in the OSNMA message?
        \begin{enumerate}
            \item If it stands as a limitation, is over-the-air rekeying (OTAR) expendable?
        \end{enumerate}
        \item What is the maximum size available for tags/signatures in the OSNMA message?
    \end{enumerate}
    \item Would it be necessary to change the design of OSNMA in a post-quantum scenario?
    \begin{enumerate}
        \item Could it be implemented in a full public-key system (e.g. leaving out TESLA)? 
    \end{enumerate}
\end{enumerate}

\subsection{Objectives}

The main objective of our work is to analyse and assess the possibilities to implement secure PQC algorithms in OSNMA and establish a framework for analysis and discussion for future Galileo System Builds and other projects with similar characteristics (e.g. satellite-based, low bandwidth, limited resources, etc.). We are aiming

\begin{enumerate}
    \item To review and document the current OSNMA design, as knowledge support for future researching.
    \item To select those OSNMA parameters that would make feasible a cryptographic agility strategy.
    \item To identify limitations of PQC implementations regarding satellite communication channels.
\end{enumerate}

\subsection{Outline}

This paper is structured as follows:

\begin{enumerate}

    \item Section \ref{sec:s2} documents the background needed to follow our study related to the Galileo system, the OSNMA service and PQC.
    \item Section \ref{sec:s3} explains the methodology designed to fulfil our objectives and answers the research questions. 
    \item In Section \ref{sec:s4} we perform an overview of the OSNMA processing logic, and the PQC algorithms that are being selected for standardization, to identify the key elements that would have an impact on the transition process.
    \item In Section \ref{sec:s5} the system's elements are analyzed, evaluating the characteristics, requisites and constraints of the OSNMA system.
    \item Section \ref{sec:s6} discusses the results, leading to the assessment of how PQC could be implemented.
    \item Section \ref{sec:s6} concludes the document by presenting the findings, and raising open questions that could lead to future lines of research.
    
\end{enumerate}

\section{BACKGROUND}
\label{sec:s2}

\subsection{Galileo}

Galileo, the European GNSS program, declared its Initial Services in December 2016. Since then, the performance of Galileo has been gradually improving thanks to the addition of satellites to the constellation, the evolution of the ground segment’s infrastructure and the deployment of new services. Upon completion, users will benefit from its full first-class performance, reliability and coverage, providing the following services (\cite{european_gnss_supervisory_authority_galileo_2021}):

\begin{itemize}
    \item Open Service (OS): ``Open and free of charge service, interoperable with its other GNSS counterparts (\cite{gaglione_benefit_2015}), for positioning and timing provision'' (\cite{european_gnss_service_centre_services_nodate}).
    \item Open Service Navigation Message Authentication (OSNMA): ``Free access service complementing the OS by delivering authenticated data, assuring users that the received Galileo navigation message is coming from the system itself and has not been modified''.
    \item Public Regulated Service (PRS): ``Service restricted to government-authorised users, for sensitive applications that require a high level of service continuity'' (\cite{eu-parliament_decision_2011}).
    \item High Accuracy Service (HAS): ``A free access service complementing the OS by delivering high accuracy data and providing better-ranging accuracy, enabling users to achieve sub-meter level positioning accuracy'' (\cite{fernandez-hernandez_galileo_2022}).
    \item Commercial Authenticated Service (CAS): ``A service complementing the OS, providing a controlled access and authentication function to users'' (\cite{noauthor_galileo_nodate}).
    \item Search and Rescue Service (SAR): ``Europe’s contribution to the international satellite-based search and rescue distress alert detection system COSPAS-SARSAT'' (\cite{zurabov_cospas-sarsat_1998}), which enhances the coverage of the system and includes a return link. 
\end{itemize}

\begin{figure}[!ht]
\centerline{\includegraphics[width=88mm]{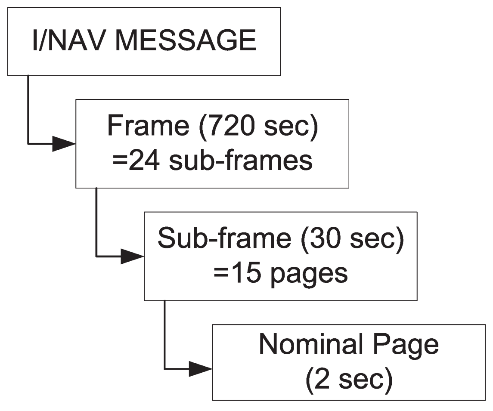}}
\caption{I/NAV Message Structure (\cite{european_gnss_supervisory_authority_european_2021})}
\label{fig:fig1}
\end{figure}

The Galileo OS navigation message I/NAV, described in the Signal in Space (SiS) Interface Control Document (ICD) (\cite{european_gnss_supervisory_authority_european_2021}), is broadcasted through the E1 (E1-B signal) and E5b (E5b-I signal) bands, and encoded in the following format (see Figure \ref{fig:fig1}):

\begin{itemize}
    \item Pages are the basic elements of the I/NAV. There are two types of Pages, even and odd ones, being transmitted every 2 seconds and comprising 120 bits each. The OSNMA information is transmitted only in the odd Pages of the E1-B signal, specifically using a 40-bit reserved field. 
    \item A Sub-frame is transmitted every 30 seconds, and it is composed of 15 Pages. Each Sub-frame contains an OSNMA complete message.
    \item The full OS message is broadcasted in Frames (decomposed in 24 subframes), every 720 seconds.
    
\end{itemize}

\subsection{Galileo OSNMA}

\begin{figure}[!ht]
\centerline{\includegraphics[width=88mm]{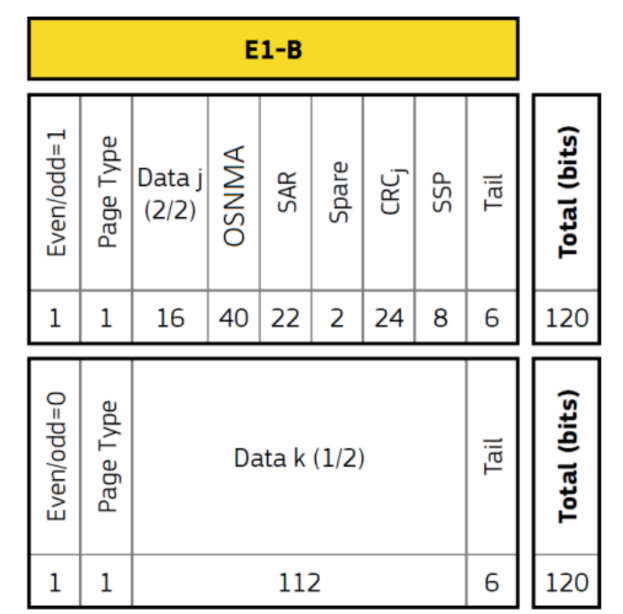}}
\caption{Allocation of odd OSNMA data within the I/NAV (\cite{european_gnss_supervisory_authority_european_2021})}
\label{fig:fig2}
\end{figure}

With OSNMA, Galileo provides an authentication mechanism that works not only for validating its own OS messages but also for other satellite navigation systems, like NAVSTAR GPS (\cite{nicola_gps_2021}). OSNMA is implemented by taking advantage of the 40-bit reserved field available in the OS I/NAV message, as it is shown in Figure \ref{fig:fig2}. This cross-validation relies on a hybrid cryptosystem (i.e. using both symmetric and asymmetric cryptography).

The key generation process in TESLA (\cite{perrig_timed_2005}) consists of a chain of subkeys derived by a one-way function. These keys feed, in reverse order, a symmetric-key-based authentication function that signs the message. The signing operation in OSNMA is performed using Hash-Based Message Authentication Codes (HMAC) (\cite{krawczyk_hmac_1997}), and the resulting message authentication code is truncated to obtain a short-length tag.

\begin{figure}[!ht]
\centerline{\includegraphics[width=150mm]{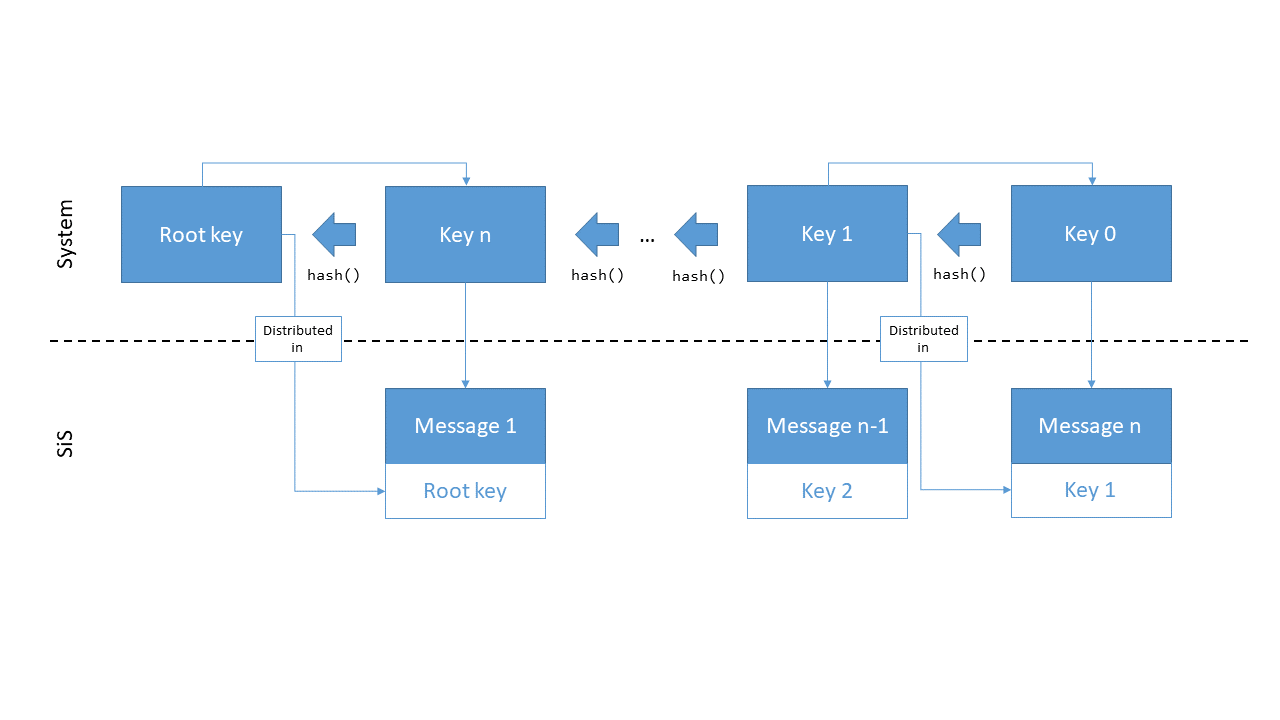}}
\caption{TESLA high level processing logic}
\label{fig:fig3}
\end{figure}

The Root key (i.e. the last element derived in the chain generation, as shown in Figure \ref{fig:fig3}, acts as a validation key that is securely distributed at the beginning of a chain period, signed with the public elliptic curve key (\cite{pornin_deterministic_2013}). This Root key is never used to sign messages. Rather, future keys shall be hashed, using the TESLA chain generation algorithm, until the user reaches the Root key again. This one-way mechanism prevents revealing any information that compromises future keys from present ones, allowing forward validation.

The elliptic curves used in the current OSNMA implementation are ECDSA P-256 and ECDSA P-521, and the corresponding public keys are verifiable through a Merkle Tree publicly available on the program webpage (\cite{european_gnss_supervisory_authority_galileo_2021-1}).

\subsection{Galileo OSNMA Message}

As described earlier, a complete authentication message is split among several OS Pages, taking advantage of a reserved 40-bit field. The OSNMA message is documented in its own SiS ICD (\cite{european_gnss_supervisory_authority_galileo_2022}), at both Page and Sub-frame levels (i.e. joining 15 Pages). This message would be broadcast by several satellites at the same time, but not by the whole constellation.

\begin{figure}[h]
    \centerline{\includegraphics[width=150mm]{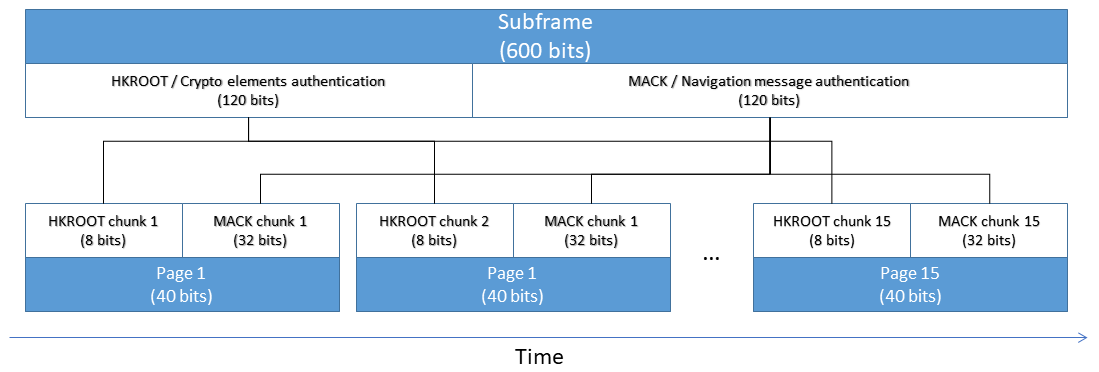}}
    \caption{OSNMA information distribution}
    \label{fig:fig4}
\end{figure}

Each Sub-frame contains two types of OSNMA information, evenly distributed throughout the Pages (as shown in Table \ref{fig:fig4}):

\begin{itemize}
    \item HKROOT: distributing the asymmetric cryptographic material needed to validate authentication information and TESLA keys.
    \item MACK: distributing the information needed to validate the I/NAV, and the previous OSNMA messages (i.e. TESLA keys).
\end{itemize}

Therefore, even though it will not be reliable until the validation keys are broadcasted 30s later, the user would retrieve almost all the information needed to authenticate the I/NAV.

\subsubsection{HKROOT (per subframe)}

Every HKROOT message contains, in turn, a Data Signature Message (DSM) Block. The DSM will distribute the Root key of the TESLA chain in force (DSM-KROOT), or the elliptic curve public keys for its authentication (DSM-PKR). In the ``NMA Header'' (see Table \ref{tab:tab1}) there are two relevant fields: ``Chain ID'' (CID), updated every time the TESLA chain in force changes; and ``Chain and Public Key Status'' (CPKS) which indicates when there is a change in any cryptographic asset.

The DSM ID field specifies which type of DSM is being broadcasted. To build the complete DSM message, several DSM Blocks must be retrieved; but as the length of the DSM Block ID parameter that is present in the HKROOT DSM Header, is 4 bits long, all the information must be allocated in 16 DSM Blocks at most.

\begin{table}
    \centering
    \begin{tabular}{cc|c|c|c|c|c|}
    \cline{3-7}
    & &  &	\multicolumn{2}{c|}{DSM Header} &	 &	 \\
    \cline{4-5}
    & & NMA Header &   DSM ID	&    DSM Block ID		 &	DSM Block n &	Total \\
    \hline
    \multicolumn{2}{|c|}{Size(bit)}  &	8 &    4 & 4 & 104 &	120 \\
    \hline
    \end{tabular}
    \caption{HKROOT Message format}
    \label{tab:tab1}
\end{table}

On one hand, the DSM-PKR message, with a length of $l_{DP}$ bits (for clarity, Appendix \ref{apx:apx1} contains a summary of the variables used in OSNMA), has two fields that are relevant to our study:

\begin{itemize}
    \item NPKT (4 bit): specifies the New Public Key Type. Only 1, 3, and 4 are assigned, so the 13 remaining values, marked as reserved in the ICD, could be used.
    \item NPK ($l_{NPK}$ bit): the actual New Public Key broadcasted.
\end{itemize}

On the other hand, the DSM-KROOT ($l_{DK}$ bit) message contains this information:

\begin{itemize}
    \item HF (2 bit): Hash Function used for building the TESLA chain, which could be either SHA-256 or SHA3-256.
    \item MF (2 bit): MAC function used for signing the navigation information, to be chosen between HMAC-SHA256 or CMAC-AES.
    \item KS (4-bit): an integer that specifies $l_K$, i.e. the length of the chain Keys. Values from 9 to 15 are reserved, so they could be used later.
    \item TS (4 bit): an integer that specifies the size of the message signatures lT (i.e. Tags). Values from 0 to 4, and from 10 to 15, are also reserved.
    \item KROOT ($l_K$ bit): Root key of the TESLA chain in force.
    \item DS ($l_{DS}$ bit): The Digital Signature of KROOT.
\end{itemize}

The election of specific algorithms and parameters applied to a message are specified using the MACLT field. This field is present in DSM-KROOT but is not documented here because of its lack of relevance to our study. The strategy specified by MACLT is known as ``Authentication Data and Key Delay'' (ADKD) and will be finally materialized into the MACK message. 

\subsubsection{MACK (per subframe)}

With the cryptographic elements provided by the DSM in our possession, now we have to validate the signatures of the navigation message. The information needed for this authentication process is broadcasted within the MACK message. As Table \ref{tab:tab2} illustrates, MACK will contain the signatures and the keys necessary to validate earlier distributed signatures ensuring, with this delay, protection against spoofing.

\begin{table}
    \centering
    \begin{tabular}{cc|c|c|c|c|c|}
        \cline{3-7}
         && Header & Tags &  Key &   Padding &   Total  \\
         \hline
         \multicolumn{2}{|c|}{Size (bit)} &   36 to 56 (depending on $lT$) &   224 to 360 (depending on $nT$) &	$l_K$	& Remaining to ``Total''	& 480 \\
         \hline
    \end{tabular}
    \caption{MACK message specification}
    \label{tab:tab2}
\end{table}

The Tags section contains nT signatures followed by some metadata. More precisely, each Tag would contain the signature (i.e. HMAC) of the part of the navigation message specified, either by KROOT’s MACLT or by a specific ADKD value set up in the Tag Info field. ADKD indicates whether ephemeris, time, or which other data is being authenticated, and should be coherent with MACLT. Moreover, the Tag contains a field (i.e. Data Cut-Off Point, COP), that indicates the elapsed time between the signature, and the data authenticated.

Finally, the Key is the element used to generate the preceding MACK message signatures. There is usually a delay of one MACK message (i.e. 30 seconds) between the Tags and the key used to generate them. However, a special dedicated ADKD strategy can also be employed, where this offset consists of 10 subframes to allow a slow validation (i.e. after 5 minutes).

\subsection{Post-quantum cryptography}
\label{sub:pqc}

Cryptographic security relies on the computational complexity of some mathematical problems. This complexity is usually expressed as the asymptotic time that a computer requires to solve it when the size of the problem grows (e.g. when increasing the number of bits of a number that must be factorized). If the problem cannot be solved in polynomial time, it serves as a proof of hardness that enables its usage for cryptographic algorithms. However, the classical complexity classes that characterize this hardness have always been linked to the classical computational paradigm and are not always resistant enough against the new quantum-based computers. 

To face this threat, there are several mathematical problems that by now have proven to present enough complexity against classical and quantum adversaries. A secure algorithm relies on the assumption that breaking its encryption mechanism is equivalent to solving one of these hard mathematical problems. In this sense, the most relevant primitives for our study are:

\begin{itemize}
    \item Lattice-based problems: the most prolific approach, in terms of the number of proposed post-quantum algorithms (\cite{computer_security_division_announcing_2022}), is based on hardness assumptions over discrete vector spaces, e.g. Shortest Vector Problem, Learning With Errors, etc. These problems are also in the spotlight due to their usefulness in the homomorphic encryption area (\cite{lyubashevsky_ideal_nodate}).
    \item Hash-based problems: they stand for algorithms supported by one-way functions, so they are especially focused on digital signatures. There are constraints related to the times the same key can be used, so in some schemes, keeping track of the operations performed is mandatory. Therefore, this family of problems is divided into two categories: stateful and stateless algorithms. The first one has been handled in a separate competition by NIST (\cite{computer_security_division_stateful_2018}), XMSS (\cite{huelsing_xmss_2018}) and LMS (\cite{mcgrew_leighton-micali_2019}) were approved as secure post-quantum algorithms. However, in contrast with some stateless proposals, the stateful approaches do not fit with performance requirements, and they haven’t been included in the PQC competition.
\end{itemize}

There are also promising families of problems based on multivariate polynomials (\cite{federal_office_for_information_security_bsi_quantum-safe_nodate}), and also on isogenies of elliptic curves (\cite{galbraith_computational_2018}). Still, none of the proposals have stood the test of time and some have even failed against classical computers (\cite{castryck_efficient_2022}).

The cryptography that implements these quantum-resistant problems, is known as Post-Quantum Cryptography (PQC), and the most acknowledged process of PQC standardization is the one that has been developed by NIST since 2017 (\cite{computer_security_division_post-quantum_2017}). We can segregate this process into two families of algorithms: Key Encryption Mechanisms (KEM) and Digital Signature Algorithms (DSA).

Even though they are still looking for new proposals (\cite{computer_security_division_round_2017}), some algorithms have already been selected for standardization (\cite{computer_security_division_selected_2017}). In this regard, the DSA algorithms accepted by NIST are:

\begin{itemize}
    \item CRYSTALS-Dilithium (\cite{bai_crystals-dilithium_2021}): together with its counterpart in KEM (i.e. CRYSTALS-Kyber (\cite{avanzi_crystals-kyber_2021})), this scheme works under the lattice-based problems hardness assumptions (\cite{lyubashevsky_fiat-shamir_2009}).
    \item Falcon (\cite{fouque_falcon_2020}): based on the well-known lattice scheme NTRU (\cite{hoffstein_ntru_1998}), it stands out for its speed and compactness.
    \item SPHINCS+ (\cite{aumasson_sphincs_2022}): a hash-based scheme, that makes use of Winternitz One Time Signatures (\cite{hulsing_wots_2017}) combined with hypertrees to obtain a stateless signature system.
\end{itemize}

One of the major challenges of these schemes, apart from the security, is the usability. Albeit the cryptographic management remains similar to the classic public key schemes, the size of crypto messages (e.g. signatures) has become larger, and it doesn't fit for most of the current applications' boundaries.

\subsection{PQC challenges}

The replacement of classical cryptographic algorithms with new quantum-resistant ones is not a trivial task. Regardless of the standardization and testing process, once an algorithm proves to be reliable, there is always the possibility of finding new vulnerabilities. The mathematical areas of the current proposals are too complex for a wide understanding, so their audit is restricted to certain people with a strong mathematical background. 

For this reason, the consensus proposes the use of hybrid approaches (\cite{stebila_hybrid_2023}) (i.e. combining classical and PQC schemes), and the implementation of a dynamic selection of algorithms in the protocols, coining the term ``crypto agility'' (\cite{ott_identifying_2019}) to refer to that.

Moreover, regardless of the initiatives for avoiding implementation failures, especially related to the election of parameters (e.g. CIRCL (\cite{faz-hernandez_introducing_2019}), Open Quantum Safe (\cite{noauthor_open_nodate}), etc.), it must be taken into account that other types of issues, such as side-channel attacks (\cite{noauthor_survey_nodate}), could come out.

However, the main challenges of the transition to PQC are related to fitting the cryptographic elements into the protocols most used today.  The sizes of the cryptographic elements proposed in PQC algorithms are significantly larger than the old ones (\cite{westerbaan_sizing_2021}), and not all the network protocols support them: e.g. the use of ``Maximum Segment Size'' or ``Maximum Transfer Unit'' fields of TCP (\cite{celi_post-quantum_2022}), the fields in X.509 public key certificates (\cite{boeyen_internet_2008}), etc. 

Computational resource consumption is another element to consider when implemented in a server, as it should attend to multiple clients’ demands in a finite time. Some researches address the implementation of PQC in the Internet’s building blocks (\cite{stebila_post-quantum_nodate}), and the most used secure protocols like TLS (\cite{stebila_hybrid_2023}) or SSH (\cite{sikeridis_assessing_2020}), validating the suitability of the PQC standardized algorithms but always focusing on endowing the protocols of crypto agility through the hybridization on cryptographic controls. From the industry side, Google and Cloudflare have carried out experiments in this regard, showing that it is feasible to implement PQC in public-testing environments (\cite{kwiatkowski_tls_2019}).

In the key exchange area, there are alternative approaches on the table, like Quantum Key Distribution (QKD) (\cite{ntanos_entanglement-based_2023, alvaro_caramuel_2022}), but nowadays the software transition is a priority (\cite{federal_office_for_information_security_bsi_quantum-safe_nodate}). Particularly in the establishment of security parameters when tunnelling a connection (e.g. IKEv2 on IPsec, one of the most enabler protocols for building secure architectures), there exist successful proposals for implementing PQC (\cite{pazienza_analysis_2022}).

\section{METHODOLOGY}
\label{sec:s3}

This assessment requires the analysis of several technologies and documents, that cover diverse areas of the OSNMA system: from the highest levels of the protocol, to the roots of the cryptographic primitives. To perform our study, the next steps will be followed:

\begin{enumerate}
    \item To develop an analysis of the relevant fields of the OSNMA message, to detect which could suffer the most impact due to the implementation of PQC. For this task, we will analyse the official documentation for both OS and OSNMA, in particular, the OSNMA SiS ICD (\cite{european_gnss_supervisory_authority_galileo_2022}). Aligned with the first research objective, this task allows to detection of the OSNMA message fields susceptible to holding the PQC material and answering research questions 1(a) and 1(b).  
    \item To review available PQC algorithms and their characteristics. Once identified which algorithms have been formally tested and are widely accepted, we will analyse the technical specifications of each of them. This process allows us to get the basic information necessary to evaluate how they fit in the previously analysed ONSMA fields, covering the second research objective. 
    \item To relate the results of the previous analysis steps to perform an assessment and discuss the key findings that lead to conclusions about how PQC can be implemented, or not, within the current design of OSNMA. It will lead to cover the third objective of the present paper and provide answers to research question 2. 
\end{enumerate}

\section{TRADE-OFFS}
\label{sec:s4}

\subsection{OSNMA characteristics}

\begin{figure}[!ht]
\centerline{\includegraphics[width=150mm]{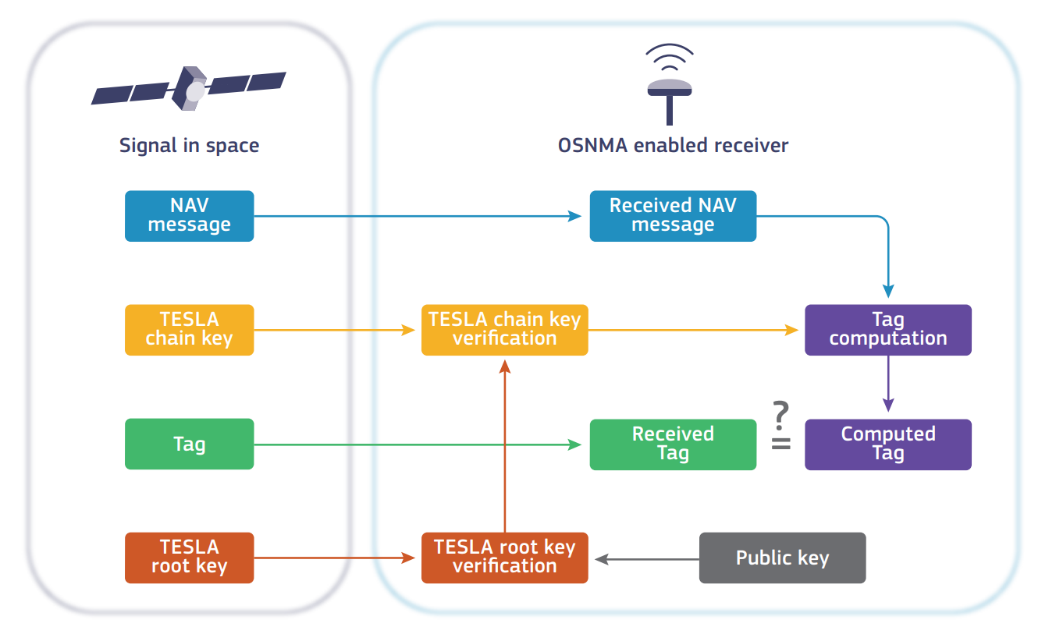}}
\caption{OSNMA processing logic (\cite{european_gnss_supervisory_authority_galileo_2022})}
\label{fig:fig5}
\end{figure}

The evaluation of the different approaches available for implementing PQC at OSNMA requires the previous analysis of the authentication lifecycles, considering the constraints that they could be subject to. The process for authenticating navigation messages has a different lifecycle (see Figure \ref{fig:fig5}) depending on the cryptographic material that is focused, namely: Merkle Tree (MT), Elliptic Curve (EC), TESLA Keychain’s Root key (KR), and TESLA Key (K).

At the beginning of the OSNMA project, 16 EC key pairs were generated. These keys that, in nominal conditions, should never be changed, act as leaves for building the MT. In the case of an incident enforces to revoke them, there is a mechanism for notifying the clients through the SiS, using a CPKS special value. 

\begin{figure}[!ht]
\centerline{\includegraphics[width=181mm]{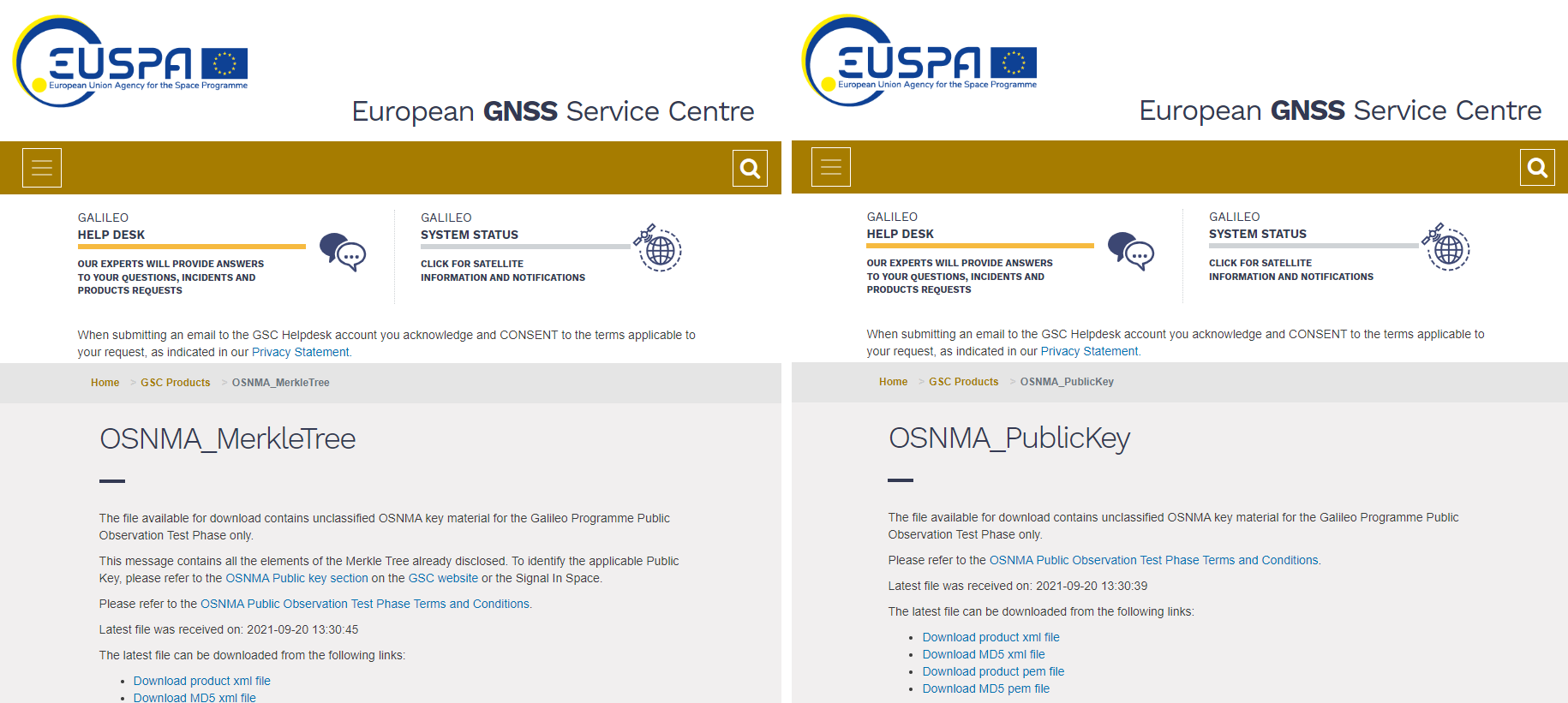}}
\caption{OSNMA portal at GSC website (\url{https://www.gsc-europa.eu})}
\label{fig:fig6}
\end{figure}

Both MT and the EC keys can be retrieved from the Galileo Service Centre (GSC) portal, as shown in Figure \ref{fig:fig6}. While the MT must be installed manually into the receivers (i.e. navigation devices), the EC keys would be also distributed through the SiS. 

When the EC keys are broadcasted over the air, into the DSM-PKR message, they can be validated using the previously downloaded MT information. This dissemination occurs every 6 hours, for a duration of 30 minutes. The EC keys can be in force for several years and when they must be updated, it is notified using the CPKS field.

Once the public key material is generated, the basis for the symmetric key authentication is set up. With a minimum period of once per day and a maximum of once per hour, a new TESLA key chain is generated and the change is also notified via CPKS. The new KR is distributed into the DSM-KROOT message, being signed with the EC key in force.

Depending on the ADKD value, a part of the navigation message is signed using HMAC (or the function specified at this moment) and the correspondent key K from the TESLA chain, as illustrated in Figure \ref{fig:fig5}. The signature is truncated to fit lT, and distributed into a MACK message. This key K, used for generating the signature is distributed in the following MACK message, that is to say, 30s later. Finally, any Ki can be verified using other Kj authenticated before (i.e. following the key chain) till KR. 

Thus, as the constraints present in ONSMA are mainly motivated by the bandwidth, at the time of analysing cryptographic changes, there are two main aspects to consider: the size of the keys, and the time they stay in the SiS (i.e. periodicity, the sum of necessary messages to encode all the data, etc.). Table \ref{tab:tab3} provides a summary of these elements to clarify the later discussion.

\begin{table}
    \centering
    \begin{tabular}{c|c|c|c|c}
         & Size & Life period & Broadcast Periodicity & Distributed in  \\
         \hline
         MT &   N/A &   Lifetime &  N/A &   GSC OSNMA Server \\
         EC &   $l_{NPK} \leq 536$ bits &   Years &  $6$ hours &   GSC OSNMA Server, DSM-PKR \\
         KR &   $l_{K} \leq 512$ bits &   1 hour - 1 day &   1 hour - 1 day &  DSM-KROOT \\
         K &   $l_{T} \leq 40$ bits &   $30$ s &  $30$ s &   MACK \\
    \end{tabular}
    \caption{OSNMA constraints}
    \label{tab:tab3}
\end{table}

\subsection{Cryptographic specifications}

Depending on the desired security confidence level, among other considerations, the parameters of cryptographic primitives can be tuned. While in some cases the election of these parameters is related to mathematical properties (\cite{rivest_are_nodate, steinfeld_security_2004}) or implementation techniques (\cite{nemec_return_2017}), the last step usually moves the discussion to a trade-off between security and size.

The current system uses two different public key algorithms to sign the TESLA Root Keys, both based on elliptic curve cryptography: ``ECDSA P-256/SHA-256'' and ``ECDSA P- 521/SHA-512''. The security of these algorithms relies on two NIST-selected curves (i.e. P-256 and P-521), and depending on the election, the lengths of the signatures is 512 or 1056 bits, respectively. This signature is settled in the DS field of the DSM-KROOT message. 

On the PQC side, all the algorithms selected by NIST, and the vast majority of the proposed, suffer from the same problem: both signature and public key sizes are larger than the classic ones. The documentation of the NIST finalists specifies their parameters attending to NIST security levels (\cite{noauthor_request_2016}). For each of the previously referenced algorithms, Table \ref{tab:tab4} shows the minimum size requirements of their less demanding specification, namely: Dilith2 for the NIST’s level 2 approach of Dilithium, Falcon-512 for the Falcon level 1, and SPHINCS+-128s, for the SPHINCS+ security level 1 parameter set.

\begin{table}
    \centering
    \begin{tabular}{c|c|c|c}
         Algorithm &    Specification & Signature size (bit/B) &    Public key size (bit/B)  \\
         \hline
         Dilithium &    Dilith2 &       $19360$ / $2420$ &          $10496$ / $1312$ \\
         Falcon &    Falcon-512 &       $5328$ / $666$ &          $7176$ / $897$ \\
         SPHINCS+ &    SHAKE-128s &       $62848$ / $7856$ &          $256$ / $32$ \\
    \end{tabular}
    \caption{PQC algorithms characterization}
    \label{tab:tab4}
\end{table}

On the other side, the stateful hash-based alternatives have even larger sizes, either for the signature or the public keys. Even though this family of algorithms has been recognized by NIST as secure against quantum computers, the management complexity necessary to achieve security using them discourages its use and has led to standardizing them in an independent project (\cite{computer_security_division_stateful_2018}). Regardless of their security level, just for illustrative purposes, Table \ref{tab:tab5} presents the parameter combinations that achieve shorter signatures and keys in XMSS and LMS schemes.

\subsection{Synthesis recapitulation}
\label{sub:recap}

\begin{figure}[!ht]
\centerline{\includegraphics[width=181mm]{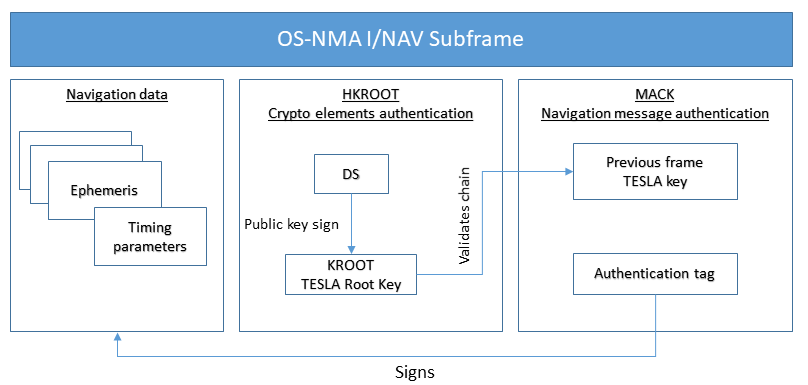}}
\caption{Block structure of OSNMA cryptographic elements}
\label{fig:fig7}
\end{figure}

With Figure \ref{fig:fig7} we aim to provide the reader with a high-level view of how ONSMA operates. It describes the cryptographic relation between the most important elements of the navigation message authentication: the navigation data, the HMAC tag that authenticates this data, the TESLA key used to generate HMAC tags, the TESLA Root Key that validates all the TESLA chain, and the Public Key used to authenticate the TESLA Root Key. A security breach in the I/NAV authentication protocol can be feasible by three key points:

\begin{enumerate}

    \item Authentication tag
    
    The navigation message is distributed with an authentication tag, which allows to verify the navigation data has been generated in a genuine satellite. A vulnerability in the function used to sign the navigation data, both in the implementation and the relying hash function, would lead to a collision. The current implementation relies on the fact that a user will get the signed navigation message before an adversary gets the signing key so that a spoofing attack can be detected. However, if these tags are not long enough, an exhaustive attack (i.e. sending a faked message with several tags until a collision is found) may be possible. It does not protect, also, against meaconing, or offline recordings.
    
    \item TESLA Chain
    
    Each authentication tag is generated hashing navigation data with a key. This key is not distributed along with the tag but in a later message. As Figure \ref{fig:fig3} shows, a user can verify that any key is part of the current TESLA chain, using the TESLA derivation function well until a previously known key is found, or the proper TESLA Root Key. A vulnerability in the TESLA chain generation would imply a vulnerability of the hash function in use.

    \item Public key cryptography
    
    Finally, TESLA bootstrapping authenticates the Root Key using public key cryptography in two steps: 1. an elliptic curve public key authenticates the TESLA Root Key, and 2. this public key is authenticated by a Merkle tree. It implies that a failure in the public key cryptography part would imply breaking the whole authentication process. An eventual quantum or classical computing vulnerability in public key cryptography would make feasible the distribution of false public keys (i.e. if the vulnerability was found in the Merkle tree cryptography) or would give an adversary access to the elliptic curve private keys, letting him create false TESLA key chains.

\end{enumerate}

\section{ANALYSIS}
\label{sec:s5}

Even though there is still room for scientific contributions (\cite{hosoyamada_cryptanalysis_2018}), the quantum algorithms available for breaking the security of symmetric key schemes do not represent a threat nowadays. Merkle Tree cryptography relies on hash-based principles, so it is also safe against quantum computers (\cite{buchmann_merkle_2008}). However, elliptic curves would be deeply vulnerable against Shor's algorithm, becoming the link that would risk the whole authentication chain. If an adversary had access to a stable quantum computer with enough resources to implement Shor’s algorithm he could retrieve OSNMA private keys from the public ones. There are also novel research paths approaching the algorithm's implementation over noisy equipment (\cite{gidney_how_2021}). This would lead him to forge a navigation message that would be valid in terms of authentication, placing the users in a spoofing risk scenario.

Regarding the taxonomy proposed by Celi (\cite{celi_post-quantum_2022}), the main challenge at OSNMA is the limited available bandwidth and the implications it would have if a user had to wait for a larger key or signature. Additionally, a major change in the protocol or the fields of the message would lead to leaving inoperative many critical or embedded devices that make use of OSNMA but cannot be updated easily. So any proposed upgrade has to fit with the current configuration of the SiS.

Finally, in the absence of specific requirements, the computational overload of the cryptographic operations is not a major concern for the performance of the system. The main issue in the protocols documented earlier (e.g. TLS, SSH, etc.) is the role of the servers as meeting points of several clients at the same time, but the satellites’ workload does not depend on the number of receivers.

Regarding cryptographic agility, the use of new cryptographic methods raises additional risks. As it is discussed in (\cite{fernandez-hernandez_pppppp-rtk_2023}), vulnerabilities due to the lack of testing or even failures in the implementation can emerge over these novel approaches. Through cryptographic agility, these risks can be mitigated. It can be accomplished by designing the systems so they can switch dynamically the set of algorithms and cryptographic primitives in use. To achieve this goal in OSNMA, the field NPKT at DSM could be used, assigning the currently free values (i.e. 0, 2 and from 5 to 15) to different combinations of algorithms.

The most popular approach to achieve cryptographic agility is hybridization (i.e. combining and overlapping classic and quantum-resistant algorithms), but the key point in OSNMA is the optimization of the bandwidth. The constraint here is that any proposed algorithm must be characterized and mapped with the currently available fields of the message. For this reason, the size of the new public keys is relevant, but the size of the TESLA Root Keys signatures is even more worthy of attention. While the public keys are rarely updated, and ultimately they can be sent by other means (e.g. over the Internet), the TESLA Root Keys are updated frequently.

\begin{figure}[!ht]
\centerline{\includegraphics[width=150mm]{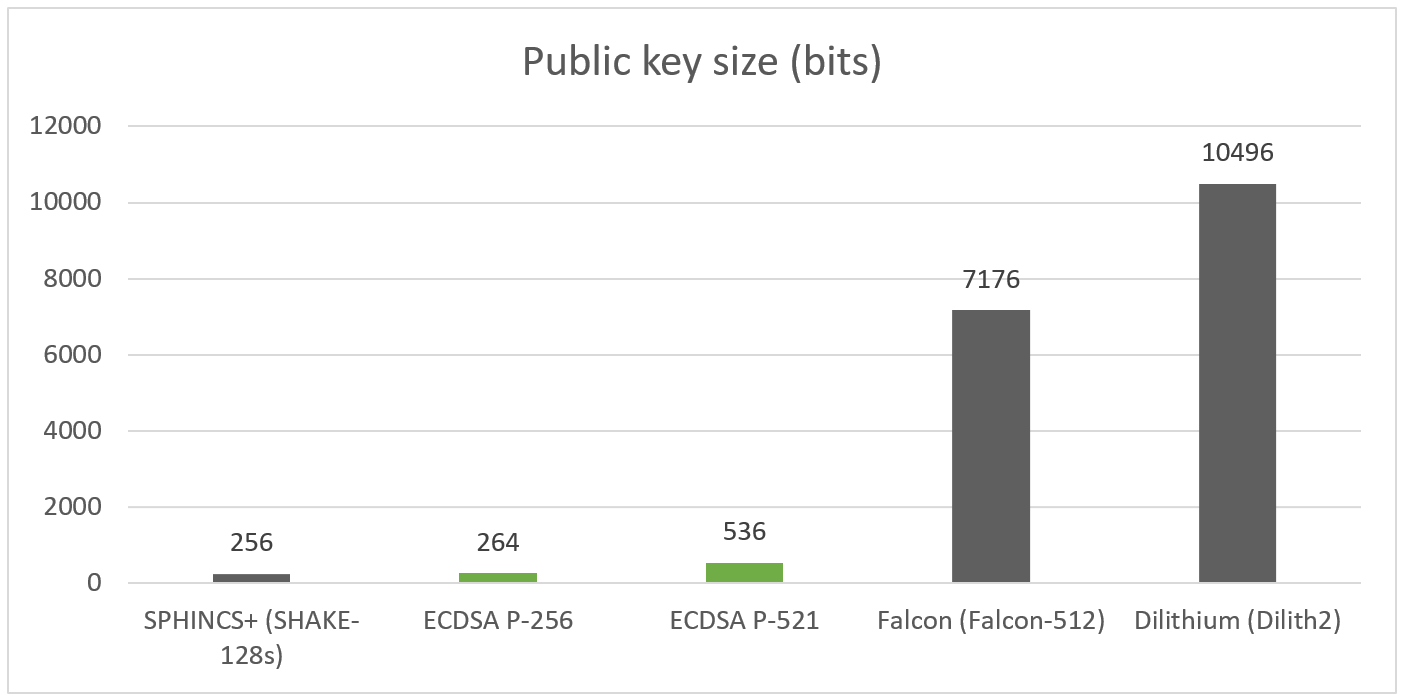}}
\caption{Size of the public keys (bits/algorithm). Grey columns represent PQC elements, green columns are the current elliptic curve ones.}
\label{fig:fig8}
\end{figure}

As Figure \ref{fig:fig8} shows, in contrast with the classical approaches (e.g. the RSA larger approach recommended by NIST is 3072-bit length (\cite{barker_recommendation_2015})), PQC keys are quite large. Pondering between key and signature sizes, the average smallest PQC implementation is Falcon and almost doubles the key size compared to its EC predecessors. Besides that, SPHINCS+ keys are even shorter than any other, but it has an issue with signatures.

\begin{figure}[!ht]
\centerline{\includegraphics[width=150mm]{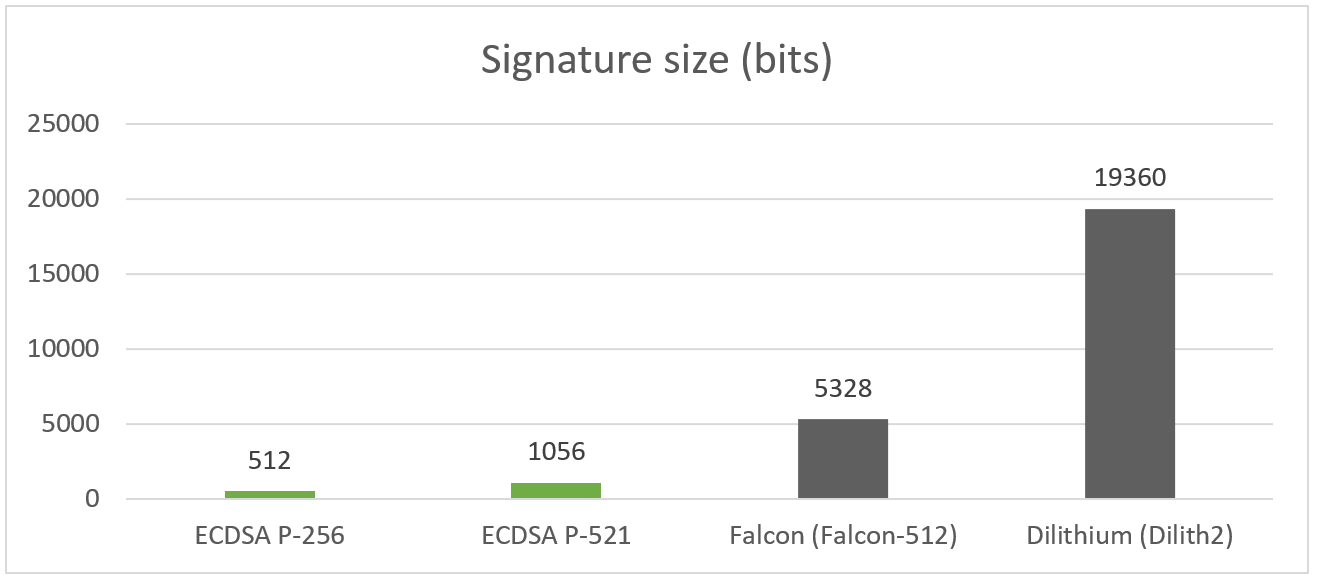}}
\caption{Size of the signatures (bits/algorithm). Grey columns represent PQC elements, green columns are the current elliptic curve ones.}
\label{fig:fig9}
\end{figure}

As previously exposed in Table \ref{tab:tab4}, SPHINCS+ signatures are around $238$ times larger than ECDSA P-264 ones so, for clarity, they haven’t been included in Figure \ref{fig:fig9}; where both EC and PQC signatures are included. Regarding the signature, Falcon is again the best quantum-resistant approach in size terms, yet it is nearly $5$ times bigger than ECDSA P-521. Moreover, the Falcon keys exceed the ECDSA P-521 by approximately $13$ points to one.

\section{DISCUSSION}

Considering the OSNMA design, there is the option of taking advantage of the 13 free values of the NPKT field, of the DSM-PKR message, to implement several authentication schemes. Therefore, it is unnecessary to modify the protocol to provide cryptographic agility.

A cryptographic failure could lead an adversary to take control of the users’ confidence in the genuine system: even though the presence of the Merkle Tree signature ensures the authenticity of EC keys if these keys were broken with future quantum techniques, a new fake TESLA chain could be generated and loaded in the users’ devices through the DSM-KROOT message, at least for an hour.

Besides these two general findings, two specific findings have also been obtained as explained below.

\subsection{TESLA implementation drawbacks}

The MACK message contains several Tags that authenticate different parts of the message and even cross-authenticate other GNSS systems’ information. As the bandwidth is limited, sending more than one tag has the following drawbacks: a) it enforces the truncation of the Tag, weakening the signature; and, b) it limits the likelihood to implement other approaches (e.g. public-key-based signatures). Especially, as argued in Section \ref{sub:recap}, the possibility of setting tag lengths to 20-bit raises many concerns about the feasibility of an exhaustive search attack (i.e. brute-forcing until finding a valid tag for a false message).

Regarding quantum, the implementation of Grover's algorithm could weaken the hash algorithms, allowing the retrieval of the complete TESLA key chain from the Root Key. Nevertheless, this algorithm is not as effective against symmetric cryptography as Shor's is against the asymmetric one, so doubling the keys' length would neutralize quantum adversaries.

If the space reserved for the several tags were just for one public key signature, it would be less versatile but more secure, and the message could be self-authenticated without depending on information from a future message. There are several fields, such as HF or MF of the DSM-KROOT message, whose existence wouldn’t be necessary if the full system were implemented using public key cryptography (i.e. without TESLA). Any change in the protocol should take this fact into account as it would improve the bandwidth.

In addition, it must be noted that the open distribution of TESLA keys makes them inappropriate for other long-term uses not related to live navigation. For instance, any process based on analyzing a signal recording (e.g. to authenticate the logs recorded in a digital tachograph) would be vulnerable to a forged signal with false navigation data.

\subsection{Post-quantum cryptography implementation}

Given the criticism of public key cryptography in the OSNMA protocol, analysed in Section \ref{sub:recap}, and the presence of algorithms susceptible to being broken by quantum computers (e.g. elliptic curve), the PQC transition should be prioritised. According to literature \cite{sosnowski_performance_2023}, as performance should not fall, we focus on bandwidth.

The fourth round of the NIST PQC competition already beholds the evaluation of KEM algorithms, so it is not foreseeable the standardisation of a new quantum-resistant digital signature algorithm in the short term. Therefore, even while NIST is still considering other algorithms, the characterization documented in Section \ref{sub:pqc} will remain valid.

Stateful hash-based algorithms are not valid for the system due to signature size reasons. It would be possible to use the XMSS configuration documented in Table \ref{tab:tab5} for the signature of the messages, as it would be just 4 times longer than the current EC signatures. However, as a stateful hash-based algorithm, its public keys must be updated frequently, and the time to transfer that GB over the air would be completely unacceptable.

\begin{table}
    \centering
    \begin{tabular}{c|c|c|c}
         Algorithm &    Parameters & Signature size (bit/B) &    Public key size (bit/B)  \\
         \hline
         XMSS &    $w=32$, $h=8$ &       $2560$ / $320$ &          $256*10^9$ / $2^{35}$ \\
         LMS &    $h=512$ &       $131072$ / $16384$ &          $2^{12}$ / $2^{9}$ 
    \end{tabular}
    \caption{Stateful Hash-based algorithms characteristics}
    \label{tab:tab5}
\end{table}

\subsection{Assessment conclusion}

In conclusion, the most suitable algorithm to replace EC would be Falcon. However, its elements do not meet the requirements to be transmitted over the air, neither for the distribution of the public keys nor for the broadcasting of the signatures. Starting from the assumption that the 4-bit field reserved for DSM Block ID (BID only admits the transmission of 16 DSM Blocks, and each one is 104 bits long, we have 1664 bits available for these two use cases:

\begin{itemize}

    \item Use case 1: transmission of New Public Keys in DSM-PKR
    
    Discarding the 16 bits of metadata, there are available 1632 bits for cryptographic material. As the Merkle Signature fills 1024 bits, there are 608 free bits for the public key (i.e. 32 bits per DSM Block). Nowadays, the Merkle Signature is the lightest quantum-resistant signature that can be implemented. Furthermore, if we were dispensed with it, even though we would save 10 DSM Blocks, we would lose the possibility to authenticate the new PK. The shortest Falcon public keys take 7176 bits, so they would need 71 DSM Blocks to be transmitted.
    
    \item Use case 2: signature of TESLA Root Key in DSM-KROOT
    
    Using the larger TESLA keys, of 256-bit length, and discarding the 104 bits of metadata, the maximum space available for the DS field (i.e. the one that holds the TESLA Root Key signature) would be 1727 bits, 79 bits per DSM Block. As the shortest Falcon signature is 5328 bits long, there would be necessary 67 DSM Blocks to cover the full authentication. 

\end{itemize}

To increase the size of the PK (i.e. $l_{NPK}$) it would be necessary to increase the number of blocks that comprise the DSM-PKR, which would also impact the bits necessary to identify these blocks at the DSM Header. It occurs the same with the signatures of DSM-KROOT. The only feasible solution would be to enlarge BID to 7 bits, getting the possibility to send up to 13312-bit long DSM messages, so the 4 necessary bits to extend BID could be subtracted from the DSM Block itself. However, it would imply that the transmission of the complete message would last 142s instead of 30s, but, as this transmission is performed at the DSM message, it will not have any overload for the authentication of the navigation data, transmitted through the MACK message.

In any case, the transmission of the digital signatures (i.e. DS at DSM-KROOT), must be prioritised over the public keys themselves (i.e. NPK at DSM-PKR). The TESLA digital signatures are sent very often, while the public keys do not usually change, and can be updated using out-of-band channels, like the Internet. However, in a full PQC approach, i.e. where the navigation message is signed using PKC instead of TESLA, it would also have an impact on the MACK message, making it longer.

In conclusion, the only quantum-resistant approach feasible nowadays, without modifying the OSNMA SiS specification, is the authentication of the message out of band. The transmission of large signatures or public keys (i.e. belonging to the documented PQC algorithms) could be performed through the Internet, or alternative channels like is done at the present with the Merkle tree, so there could exist a reliable source for validating the navigation messages. Receivers would perform the authentication as before, but using the new PQC schemas. Even though it would not be implemented in disconnected devices, it could open the door to mitigate cryptographic risks in the connected ones.

\section{CONCLUSION AND FUTURE WORK}
\label{sec:s6}

In this paper an analysis of the OSNMA authentication procedures, to provide an assessment of how to make it quantum-resistant, has been developed. To achieve it, we have analysed both the OSNMA specifications as well as the state-of-the-art about post-quantum cryptography; obtaining as an indirect contribution a well-documented summary of the whole OSNMA authentication process from different perspectives. This analysis has led to findings related to weaknesses of the system, like the ones related to the TESLA implementation.

Our study reveals that the current OSNMA design, even though it has elements that could enable cryptographic agility, is not ready for implementing the PQC algorithms that are currently available. One of the most critical points is the size of the new keys and signatures, which would slow down key distribution, but not the authentication process, as it is performed using TESLA. However, the data fields available in the current system do not have enough capacity (\cite{noauthor_shannon_nodate}) to cover the number of messages necessary to broadcast these new elements. Although it could be done with some slight changes, these modifications might be highly disruptive for the current requirements of Galileo and would thus be unfeasible.

Nevertheless, until major changes can be performed in the system, there exist alternatives to mitigate the quantum risks. For instance, some of the signatures could be delivered to the receivers by an alternative channel (e.g. over the Internet). In addition, though it is not likely to take into account the state of standardization processes, there could appear new PQC algorithms more efficient in size terms shortly. 

In any case, a transition to a quantum-resistant design has to be performed. Not doing so would imply in a few years, as stated by the National Quantum Programs (\cite{petrenko_security_2021}), an unacceptable risk for the system, and the de facto nullification of OSNMA.

This research opens the door not only to new research about PQC implementation but also to prevent some of the detected flaws in the current design:

\begin{itemize}
    \item Any change in the PQC state of the art could provide opportunities to implement a quantum-resistant scheme following the characterization performed in our study. Furthermore, the assessment of algorithms that are not covered by the NIST competition, or that have been dismissed for reasons not related to size or security requirements, could also provide valid alternatives.
    \item Although the TESLA approach is a clever manner to overcome the Galileo bandwidth limitations, as it has been already exposed, it introduces weaknesses in the system, and the revelation of the signing key can be a barrier for use cases where a long-term authentication is necessary (e.g. when authenticated audit logs must be preserved (\cite{baldini_regulated_2018})). Therefore, the replacement by a full quantum-resistant public key system should be assessed.
    \item Some of the OSNMA configurations seem to be insufficient to provide a robust authentication (e.g. the 20-bit Tags could be brute-forced). Research in this direction could help to clarify the level of risk that it implies. 
\end{itemize}

\printbibliography

\newpage
\appendix
\appendixpage

\section{OSNMA variables}
\label{apx:apx1}

Table \ref{tab:tab6} contains a summary of the OSNMA variables used within the article.

\begin{table}[!ht]
    \centering
    \begin{tabular}{c|c|c}
         Variable &     Description &                       Value \\
         \hline
         $l_{DP}$ &     Length of DSM-PKR (bit) &           $1352 \leq 104 * \lceil \frac{1040+ l_{NPK}}{104} \rceil \leq 1664$ \\
         $l_{NPK}$ &    Length of the Public Keys &         According to assigned values, from $264$ to $536$ bit \\
         $l_{PDP}$ &    Length of PKR padding &             - \\
         $l_{DK}$ &     Length of DSM-KROOT (bit) &         $728 \leq 104 * \lceil 1 + \frac{l_{K}+l_{DS}}{104} \rceil \leq 1456$ \\
         $l_K$ &        Length of the TESLA Keys &          According to assigned values, from $96$ to $256$ bit \\
         $l_T$ &        Length of the signatures/tags &     According to assigned values, from $20$ to $40$ bit \\
         $l_{DS}$ &     Length of the digital signatures &  According to assigned values, $512$ or $1056$ bit \\
         $l_{PDK}$ &    Length of KROOT padding &           -\\
         $n_{t}$ &      Number of tags per MACK message &    $4 \leq \lfloor \frac{480 - l_{K}}{l_T + 16} \rceil \leq 10$
    \end{tabular}
    \caption{Summary of OSNMA variables}
    \label{tab:tab6}
\end{table}

\end{document}